\documentclass[doublecol]{epl2}

\usepackage{amssymb}
\usepackage{amsmath}
\usepackage{graphicx}
\newcommand{\be}{\begin{equation}}
\newcommand{\ee}{\end{equation}}
\newcommand{\ben}{\begin{eqnarray}}
\newcommand{\een}{\end{eqnarray}}

\title{Phase transition in the 3-D massive Gross-Neveu model}

\author{F.C. Khanna\inst{1}, A. P. C. Malbouisson\inst{2} 
J. M. C. Malbouisson\inst{3} and A.E. Santana\inst{4}}

\institute{\inst{1}Department of Physics, Theoretical Physics Institute, University of
Alberta, Edmonton, Alberta, T6G 2J1, Canada 
and TRIUMF, 4004, Westbrook mall, Vancouver, British Columbia V6T 2A3,
Canada\\
\inst{2}Centro Brasileiro de Pesquisas F{\'i}sicas, MCT,
22290-180, Rio de Janeiro, RJ, Brazil\\
\inst{3}Instituto de F\'{i}sica, Universidade Federal da
Bahia, 40210-340, Salvador, BA, Brazil\\
\inst{4}Instituto de F\'isica, International Center for Condensed Matter Physics,
Universidade de Bras\'ilia, 70.910-900,
Brasilia, DF, Brazil}

\pacs{11.10.Kk}{Field theories in dimensions other than four}
\pacs{11.30.Qc}{Phase transition}
\pacs{11.10.Wx}{Finite-temperature field theory}

\abstract{We consider the $3$-dimensional massive Gross-Neveu
model at finite temperature as an effective theory for strong interactions. Using the Matsubara imaginary time
formalism, we derive a closed form for the renormalized
$T$-dependent four-point function. This gives a singularity,
suggesting a phase transition. Considering the free energy 
 we obtain the $T$-dependent mass, which goes to zero for some temperature. These results lead
us to the conclusion that there is a second-order phase
transition.}

\begin{document}

\maketitle

\section{Introduction}
The Gross-Neveu (GN) model~\cite{gn1}, proposed
as an effective model for quantum chromodynamics (QCD), has been analyzed
extensively in recent years. In several cases, these studies
consider the temperature effect, associated with fermion systems, in
both particle and condensed matter
physics~\cite{gn2,gn3,gn5,gn6,gn7,gn8,gn9,gn10,gn11,gn12,gn13,gn14,gn15,
gn16,gn17,gn18,gn19}. The motivations and the nature of such
analysis are multiple, including investigations of the continuous
and discrete chiral symmetry~\cite{gn7,gn8}. The version with $N$
massless fermions in (2+1)-dimensions, for instance, presents
chirality-breaking in perturbative analysis, and its restoration
at finite temperature~\cite{gn15}. This provides insight into the
intricate structure of the hadronic matter, such as for the quark
confinement/deconfinement transition~\cite{gn6,gn16,gn17,gn18}.
Due to such characteristics, the GN model is taken as a prototype
for analysis of phase transitions in quantum field theory for
fermions. In this sense, the search for analytical results is
another aspect that attracts interest, in general~\cite{gn19}, but
in particular for $N=1$, where lattice calculations are hard to
carry out. 
In this note, we  address the
(2+1)-dimensional GN model at finite temperature, but with
non-zero mass and one type of fermion. 

The massive GN model in (2+1) dimensions 
 was considered in~\cite{Jarrao}, even though this version for dimensions
greater than two is perturbatively non-renormalizable. This is justified by considering that,
perturbative renormalizability is not an absolute criterion for an
effective theory to be physically consistent. This fact is known
and has been used for
long~\cite{Gaw1,Gaw2,Weinberg1,parisi,anp,rwp1}. 
Here we will consider the (2+1)-dimensional massive GN model at 
finite temperature with an arbitrary
zero-temperature coupling constant, as an effective theory for QCD. We will take  both the thermal mass and the thermal coupling constant 
at one-loop order. This approximation will be 
justified by a non-perturbative analysis of the four-point function,
by summing the chains of loop diagrams. This provides a non-perturbative
relationship between the four-point function and temperature. Such a
relation leads to a singularity, suggesting a possible phase
transition~\cite{supc1}. In order to get the nature of the phase transition, an
analysis of the free energy is carried out. It
gives rise to a $T$-dependent mass that goes to zero for some
values of temperature, suggesting a second-order phase transition. An aditional argument in favour of a second-order phase transition comes from an analysis of the $beta$-function. The    argument relies on the fact that this analysis gives the result that 
there is a non-trivial infrared stable fixed point.

Such results bring up an interesting question: can such a study of
an effective model yield any information about a phase transition
in physically relevant situations for QCD? In order to get an
answer to this question, we rely
 on a procedure that follows methods first developed for the
BCS model in superconductivity~\cite{supc1}. 

The paper is organized in the following way. In Section 2, the
model is defined. In Section 3, the thermal correction of the mass
at the one-loop level is calculated. In Section 4, using the 
temperature-dependent coupling constant,  
 an expression for the thermal 
mass is derived. In Section 5, starting with non-perturbative
results for the four-point function, the analysis of the critical
region is discussed. The concluding remarks are presented in
Section 6.
\section{The model}
In a $D$-dimensional Euclidian manifold,
${\mathbb{R}}^{D}$, we consider the Hamiltonian for the massive GN
model,
\begin{eqnarray}
H &=&\int d^{3}x\left\{ \psi^{\dagger}(x)(\gamma ^{j}(i\partial _{j}))\psi
(x)
 -m_0\psi^{\dagger}(x)\psi (x)\right. \nonumber \\
&& \left. + \frac{\lambda_{0}}{2}\left[
\psi^{\dagger}(x)\psi (x)\right] ^{2}\right\}, 
\label{GN}
\end{eqnarray}
where $m_{0}$ and $\lambda_{0}$ are respectively the physical
zero-temperature mass and coupling constant, $x\in {\mathbb{R}
}^{D}$ and the $\gamma $-matrices are elements of the Clifford
algebra (natural units $\hbar =c=k_B=1$ are used.). This Hamiltonian
is obtained using conventions for Euclidian field theories
in~\cite{Ramond}. 

From Eq.~(\ref{GN}), introducing the thermally
corrected mass,
\begin{equation}
m(T)=m_{0}+\Sigma (T),   \label{cri1}
\end{equation}
we have a free energy density of the Ginzburg-Landau
type~\cite{linde1},
\begin{equation}
{\mathcal{F}} = a+b(T)\phi ^2(x)+c\,\phi ^4(x)
\label{july092}
\end{equation}
where  $b(T)=-m(T)$ and
$c=\lambda_{0}/2$. The minus sign for the mass in
Eq.~(\ref{july092}) implies that, in the disordered phase we have
$m(T)<0$ and for the ordered phase $m(T)>0$, consistently. The
second order phase transition occurs at the temperature where $m(T)$
changes sign from negative to positive, characterizing a spontaneous symmetry breaking. In this formalism, the quantity $\phi(x)=\sqrt{\psi^{\dagger}(x)\psi (x)} $ plays the role of the order parameter  
for the transition. 

\section{Thermal self-energy}

We evaluate the thermal self energy, $\Sigma (T),$ by using  the
Matsubara imaginary-time formalism. The Cartesian coordinates are
specified by $x=(x^{0}=\tau ,\mathbf{x})$, where $\mathbf{x}$ is a
$ (D-1)$-dimensional vector. The conjugate momentum of $x$ is
denoted by $k=(k_{0},\mathbf{k})$, $\mathbf{k}$ being a $(D-1)$
-dimensional vector in momentum space. The KMS (Kubo, Martin,
Schwinger) condition implies that the Feynman rules are modified
by the well-known Matsubara prescription,~\cite{livro}
\begin{equation}
\int \frac{dk_{0}}{2\pi }\rightarrow \frac{1}{\beta }\sum_{n=-\infty
}^{+\infty },\;\;\;\;k_{0}\rightarrow \frac{2\pi
(n+\frac{1}{2})}{\beta } \equiv \omega _{n}\,,  \label{Matsubara}
\end{equation}
where $\omega _{n}$ are Matsubara frequencies and $\beta =1/T$, $T$ being
the temperature.

At the one-loop level the finite-temperature self-energy is given by
\begin{equation*}
\Sigma (D,T,s)=\lambda_{0} \frac{m_{0}}{\beta }\sum_{n=-\infty
}^{\infty }\int \frac{d^{D-1}k}{(2\pi
)^{D-1}}\frac{1}{(\mathbf{k}^{2}+\omega _{n}^{2}+m_{0}^{2})^{s}}.
\end{equation*}
In order to use the dimensional regularization procedure, we
introduce  dimensionless quantities,
\begin{equation}
q_{j}=k_{j}/2\pi m_{0}\,,\,\, j=1,2,...,D-1 \,\,{\rm and} \,\,
a=(m_{0}\beta )^{-2} . \label{jun31}
\end{equation}
 For the present case, we have $D=3$ and
$s=1.$ After dimensional regularization~\cite{Ramond}, we obtain
\begin{equation}
\Sigma (D,T,s)=\frac{m_{0}\lambda_{0}\Gamma (\nu )}{(4\pi
)^{(D-1)/2}\Gamma (s)\beta }\sum_{n=-\infty
}^{\infty }(\omega _{n}^{2}+m_{0}^{2})^{-\nu },
\label{cri3}
\end{equation}
where $\nu =s-(D-1)/2$. The sum in Eq.~(\ref{cri3}) is cast in the
general form
\begin{equation}
\sum_{n=-\infty }^{+\infty }\left[ a(n+\frac{1}{2})^{2}+c^{2}\right] ^{-\nu
}=4^{\nu }Z_{1}^{4c^{2}}(\nu ,a)-Z_{1}^{c^{2}}(\nu ,a),  \label{EF}
\end{equation}%
where
\begin{equation}
Z_{1}^{b^{2}}(\nu ,a)=\sum_{n=-\infty }^{+\infty }\left[ an^{2}+b^{2}\right]
^{-\nu }\   \label{Es1}
\end{equation}
is the generalized Epstein $zeta$-function~\cite{ep1} defined for
$ \text{Re}(\nu)>1/2$. We then analytically continue
$Z_{1}^{b^{2}}(\nu ,a)$ to the whole complex $\nu
$-plane~\cite{ep2,jm1,jm2,jm3}.
\begin{eqnarray}
Z_{1}^{b^{2}}(\nu ,a)& =&\frac{\sqrt{\pi }}{\sqrt{a}\,\Gamma (\nu )}\left[
\frac{\Gamma (\nu -1/2)}{b^{2\nu -1}}\right. \nonumber \\
&& \left. +\,4\sum_{n=1}^{\infty }\left( \frac{\pi
n}{b\sqrt{a}}\right) ^{\nu -\frac{1}{2}}K_{\nu -\frac{1}{2}}\left(
\frac{2\pi bn}{\sqrt{a}}\right)
\right] ,
  \label{sep13}
\end{eqnarray}
where $K_{\alpha }(x)$ is the Bessel function of the third kind. As
a consequence,  $\Sigma (D,T,s)$ reads
\begin{equation}
\Sigma (D,T,s) =\widetilde{\Sigma }(D,s)\frac{m_{0}^{1-2\nu }}{\beta (2\pi
)^{2\nu }}
\lbrack 4^{\nu }Z_{1}^{4c^{2}}(\nu ,a)-Z_{1}^{c^{2}}(\nu ,a)]
\label{cri4}
\end{equation}
where $c=1/2\pi $.  Using Eq.~(\ref{sep13}) in  Eq.~(\ref{cri4}),
we obtain $\Sigma (D,T,s)$ as
\begin{equation}
\Sigma (D,T,s)=F_0(D,s)+F_1(D,a,s).\label{cri8}
\end{equation}%
Here $F_0(D,s)$ is a term independent of $a$, arising from the first
term in brackets in Eq.~(\ref{sep13}), while $F_1(D,a,s)$ is the
term arising from the sum in the second term. We find  that, for
$s=1$ and even dimensions $D\geq 2$, $F_0(D,s)$ is divergent due to
the pole of the $ \Gamma $-function. Accordingly, this term is
subtracted to get the physical a-dependent function $\Sigma
_{R}(D,a)$. The mass counter-term is a pole appearing at the
physical value $s=1$. The $a$-dependent correction to the mass is
proportional to the regular part of the analytical extension of the
Epstein $ zeta$-function in the neighborhood of the pole at $s=1$.
For  uniformity, the term $F_0(D)$ is also subtracted in the case of
odd dimensions $D$, where no poles of $\Gamma $-functions are
present. Therefore, using  the modified Bessel function for
$D=3,s=1$ and $a=(m_{0}/T)^{-2}$, i.e. $K_{\pm 1/2}(z)=\sqrt{\pi
}e^{-z}/\sqrt{2z}$,  the physical thermal self-energy  for $D=3$ is
given by
\begin{equation}
\Sigma _{R}(3,T,1)=\frac{1}{2\pi }\lambda_{0}
m_{0}^{2}f(T),
\end{equation}
where
\begin{equation}
f(T)=-\frac{T}{m_{0}}\ln \left( 1+e^{-m_{0}/T}\right),   \label{f}
\end{equation}
The behavior of $f(T)$ is the following: for $T\rightarrow 0$,$
\,f(T)\rightarrow 0$; for $T\rightarrow \infty $, $\,f(T)\rightarrow
-\infty $, with $f(T)<0$ for all values of $T>0$. Using Eq.
(\ref{cri1}), the temperature dependent mass is
\begin{equation}
m(T)=m_{0}+\frac{1}{2\pi }\lambda_{0} m_{0}^{2}f(T).  \label{cri6}
\end{equation}%
The condition for a phase transition, $m(T)=0,$ provides a
critical temperature for each fixed value of the zero-temperature
coupling constant, $\lambda_{0}$.  This result indicates  a
second-order phase transition. However, in order to establish this
result on a firmer ground, we consider the $T$-dependent
correction to the coupling constant $\lambda_{0} $.
\section{T-dependent coupling constant}
Initially, we consider the four-point function with null external
momenta, which defines the temperature-dependent coupling
constant. It is given up to one-loop by,
\begin{equation}
\Gamma _{D}^{(4)}(\lambda_{0},\beta )\simeq \lambda_{0}[1+\lambda_{0}\Pi (D,\beta
)]\;,  \label{4-point1L}
\end{equation}%
where $\Pi (D,\beta )$ is the $\beta $-dependent one-loop polarization
diagram given by
\begin{equation}
\Pi (D,\beta )=\frac{1}{\beta }\sum_{n=-\infty }^{\infty }\int
\frac{d^{D-1}k }{(2\pi
)^{D-1}}\frac{m_{0}^{2}-(\mathbf{k}^{2}+\omega
_{n}^{2})}{(\mathbf{k} ^{2}+\omega _{n}^{2}+m_{0}^{2})^{2}}.
\label{sigma0}
\end{equation}%
Using the dimensionless quantities introduced in
Eq.~(\ref{jun31}), $\Pi (D,\beta )$ is given as
\begin{eqnarray}
\Pi (D,a)& =&\Pi (D,a,s)\vert _{s=2}
=\frac{m^{D-2(s-1)}}{(2\pi )^{2}}\,\sqrt{a} \nonumber \\  
&& \times \left[ \frac{1}{2\pi ^{2}}\,U_{D}(s;a)-U_{D}(s-1;a)\right]_{s=2},
\label{sigmaA}
\end{eqnarray}
where
\begin{eqnarray}
U_{D}(\mu ;a) &=&\pi ^{\frac{(D-1)}{2}}\,\frac{\Gamma (\mu -(D-1)/2)}{\Gamma
(\mu )} \nonumber \\
&& \sum_{n=-\infty }^{\infty }\left[
a(n+\frac{1}{2})^{2}+(2\pi )^{-2}
\right] ^{(D-1)/2)-\mu }.    \label{UD}
\end{eqnarray}
The function $U_{D}(\mu ,a)$ is extended to the whole complex $\mu $
-plane, resulting in
\begin{equation}
U_{D}(\mu ,a)=\frac{h(\mu ,D)}{\sqrt{a}}\left[ \Gamma (\mu -\frac{D}{2}%
)+4W(\mu ,a,D)\right] ,  \label{UA1}
\end{equation}
where $h(\mu ,D)=\pi ^{2\mu -D/2}/(2^{D-2\mu }\Gamma (\mu )) $ and
\begin{eqnarray}
W(\mu ,a,D)&=& 2\sum_{n=1}^{\infty }\left( \frac{\sqrt{a}}{n}\right)
^{\frac{
D}{2}-\mu }K_{\frac{D}{2}-\mu }\left( \frac{2n}{\sqrt{a}}\right) \nonumber \\
&&  -\sum_{n=1}^{\infty }\left( \frac{2\sqrt{a}}{n}\right) ^{\frac{D}{2}-\mu
}K_{\frac{D}{2}-\mu }\left( \frac{n}{\sqrt{a}}\right).   \label{W}
\end{eqnarray}
Using these results in Eq.~(\ref{sigmaA}), we obtain $\Pi (D,a,s)$
as
\begin{equation}
\Pi (D,a,s)=H(D,s)+G(D,a,s),
\label{HG}
\end{equation}
where $H(D,s)$ is a term independent of $a$, arising from the first
term in brackets in Eq.~(\ref{UA1}), while the term $G(D,a,s)$
arises from the  the $ W $-functions. As in the case of the thermal
self-energy, the physical thermal polarization is obtained by
subtracting the term $H(D,s)$ in Eq.~(\ref{HG}) and is given by,
\begin{equation}
\Pi _{R}(D,a)=\frac{m_{0}^{D-2}}{2^{D-2}\pi ^{D/2}}\left[ 2\pi
W(2,a,D)-W(1,a,D)\right] .
\label{W1}
\end{equation}
\begin{figure}[h]
\begin{center}
\includegraphics[{height=6.0cm,width=8.0cm}]{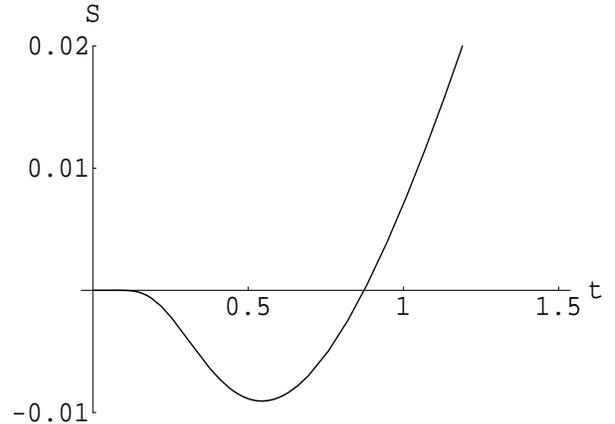}
\caption{Polarization (in units of $m_0$), $S=\Pi _{R}(3,T)/m_0$, as
a function of $T$ (in units of $m_0$, i.e. $t=T/m_0$).}
\label{figura}
\end{center}
\end{figure}

Taking $D=3$ and $a=(m_{0}/T)^{-2}$, we obtain
\begin{equation}
\Pi _{R}(3,T)=\frac{m_{0}}{2\pi }\left[ \frac{T}{m_{0}}\ln \left(
1+e^{-m_{0}/T}\right) -\frac{1}{1+e^{m_{0}/T}}\right] .  \label{sigmaR3}
\end{equation}
Substituting this expression into Eq.~(\ref{4-point1L}), we obtain
the temperature-dependent coupling constant,
\begin{equation}
g(T;\lambda_{0})\equiv\Gamma _{3R}^{(4)}(\lambda_{0},T) \simeq
\lambda_{0}[1+\lambda_{0}\Pi _{R}(3,T)]. \label{g3L}
\end{equation}
The physical polarization $\Pi_{R}(3,T)$ is shown in
Fig.~\ref{figura}. We find from this plot and from Eq.~(\ref{g3L})
that $g(T=0;\lambda_{0})=\lambda_{0}$; this just reflects the fact
that $ \lambda_{0}$  is the physical coupling constant at $T=0$.

The temperature-dependent coupling constant, $g(T;\lambda_{0})$,
is introduced in place of $\lambda_{0} $ in Eq.~(\ref{cri6}). Then
we get a $T$-dependent  mass,
\begin{equation}
m(T)=m_{0}[1+\frac{1}{2\pi }g(T;\lambda_{0})m_{0}f(T)].
\label{sep18}
\end{equation}
We have then all the elements to analyze the phase transition.

\section{The phase transition}

Let us first remind an essential feature of a situation in condensed
matter physics, theoretically close to the one we examine in this
article; namely, the BCS field theoretical approach to the
superconducting phase transition~\cite{supc1}. In this case it is
shown that, {\it at criticality}, the leading contribution to the
four-point function with zero-external momenta is given by the sum
of all chains of one-loop diagrams. This non-perturbative
calculation leads to an expression of the form
\begin{equation}
\Gamma _{3R}^{(4)}(\lambda_{0},T) =
\frac{\lambda_{0}}{[1-\lambda_{0}\Pi _{R}(3,T)]}. \label{g3L1}
\end{equation}
In our case the first two terms of the expansion in powers of
$\lambda_0$ of such a function are given in Eq.~(\ref{g3L}). The existence of 
a singularity of the four-point function in Eq.~(\ref{g3L1}) indicates a phase 
transition, as explained in~\cite{supc1}; in other words the phase transition 
is characterized by
the poles of $\Gamma _{3R}^{(4)}(\lambda_{0},T_c)$, i.e.,
\begin{equation}
\lambda_{0}\Pi _{R}(3,T)=1. \label{jun21}
\end{equation}
 In our case this singularity is found from an
analysis of Fig.~\ref{figura}. The nature of the transition is
obtained by a study of the free energy, Eq.~(\ref{july092}).

\begin{figure}[h]
\begin{center}
\includegraphics[{height=6.0cm,width=8.0cm}]{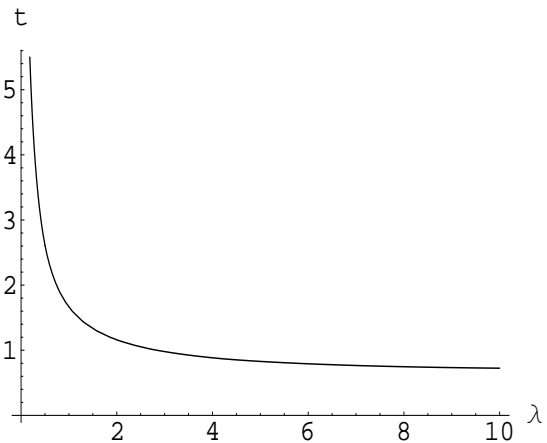}
\caption{Phase transition defined by $m(T_c)=0$: values for
the pair $(T_c,\lambda_{0})$ define the critical line; on the
vertical axis $t=T_c/m_0$ and on the horizontal axis
$\lambda=\lambda_0 m_0/2\pi$. The minimum critical temperature is
$T_{c}\simeq 0.874m_{0}$.} \label{figura3}
\end{center}
\end{figure}

In the present article we have a situation analogous to the BCS
theory, and we have done the study of the free energy for a
second-order phase transition starting from Eq.~(\ref{july092}).
Then the the critical region is defined by taking $m(T_c)=0$, in
Eq.~(\ref{sep18}), which leads to
\begin{equation}
g(T_c;\lambda_{0})=-\frac{2\pi }{m_{0}f(T_c)}.  \label{sep33}
\end{equation}
This provides $T_c$ as a function of $\lambda_0$, giving a curve
plotted in Fig.~\ref{figura3}. On the curve we have $m(T_c)=0$;
below the curve we have $m(T)>0$ and above the curve $m(T)<0$. We
find that the sign for the thermal mass corresponds to the expected
behavior of a phase transition from a disordered to an ordered
phase, as temperature is lowered.

Moreover, Eq.~(\ref{jun21}) gives rise to a function
$t=t(\lambda)$ (where again  $\lambda=\lambda_0m_0/2\pi$ and
$t=T/m_0$), with a behavior that is similar to that presented in
Fig.~{\ref{figura3}}. Indeed, from Fig.~{\ref{figura}}, we
 obtain that   $t\rightarrow  0.87$ for $\lambda_0 \rightarrow \infty$. On
 the other
hand, $t \rightarrow \infty$ for $\lambda \rightarrow 0$. This
non-perturbative result is then compatible with the analysis performed 
in the previous sections, of
the mass as a function of the temperature, taking the thermal corections to both 
the mass and the coupling constant at one-loop level. Then we conclude that
indeed  there is a second-order phase transition, characterized by the
divergence of the four point function and with the critical 
temperature obtained from the condition $m(T_c)=0$.

A further indication of a second-order transition can be obtained from a renormalization group point of view, following lines analogous to those employed in~\cite{PRB2002} for type-II superconducting films. If an infrared stable fixed point exists, it is possible to determine it by 
a study of the infrared behaviour of the 
beta-function, $i.e$, in the neighbourhood of vanishing external momentum $|p|=0$; we consider the thermal coupling constant at criticality, 
with an external small momentum, given by 
\begin{equation}
g(T, |p|\approx 0) =
\frac{\lambda_{0}}{[1-\lambda_{0}\Pi (D,T,p)]}.
\label{g3L1p}
\end{equation}
Using a Feynman parameter $x$, the one-loop four-point function is,  
\begin{eqnarray}
\Pi (D,T,p)=\frac{1}{\beta }\sum_{n=-\infty }^{\infty }\int_0^1\,dx\,\nonumber \\
 \cdot \int
\frac{d^{D-1}k }{(2\pi
)^{D-1}}\frac{M_{0}^{2}-(k^{2}+\omega
_{n}^{2})}{(k ^{2}+\omega _{n}^{2}+M_{0}^{2})^{2}}\nonumber \\
-\frac{p^2}{\beta}\sum_{n=-\infty }^{\infty }\int_0^1\,dx\, \,\int
\frac{d^{D-1}k }{(2\pi
)^{D-1}}\frac{x(1-x)}{(k ^{2}+\omega _{n}^{2}+M_{0}^{2})^{2}},
\nonumber \\
\label{sigma0p}
\end{eqnarray}
where 
\begin{equation}
M_0^2(p)=m_0^2+p^2 x(1-x)\,;\;\;\;a_c(p)=\frac{1}{\beta^2M_0(p)};
\label{Ma0}
\end{equation}
The second term in Eq.~(\ref{sigma0p}) vanishes in the limit $|p|\rightarrow 0$. Following the calculation steps for small values of $|p|$ detailed in~\cite{PRB2002}, we are left for  $|p|\approx 0$, after 
regularization, with an expression similar to Eq.~(\ref{W1}),
\begin{eqnarray}
\Pi_{R} (D,T,p)&=&\frac{M_{0}^{D-2}}{2^{D-2}\pi ^{D/2}}\left[ 2\pi
W(2,a_c(p),D)\right.\nonumber \\
&&\left. -W(1,a_c(p),D)\right] ,\nonumber \\
\label{W2}
\end{eqnarray}
where it is understood that we take asymptotic values for small values of the argument in the Bessel functions in Eq.~(\ref{W}), which defines the quantity $W(\mu ,a,D)$ . 

For $D=3$ we have, $M_{0}^{D-2}\approx m_0+\frac{1}{2}p^2x(1-x)$, and
\begin{equation}
\Pi_{R} (3,T,p)=B(3,T)+A(3,T)|p|
\label{PiAB}
\end{equation}
where,
\begin{equation}
B(3,T)=\frac{m_{0}}{2\pi ^{3/2}}\left[ 2\pi
W(2,a_c,3)-W(1,a_c,3)\right] 
\label{W3}
\end{equation}
and 
\begin{equation}
A(3,T)=\frac{a}{2\pi ^{3/2}}\left[ 2\pi
W(2,a_c,3)-W(1,a_c,3)\right], 
\label{W4}
\end{equation}
with $a=\int_0^1dx\sqrt{x(1-x)}=\pi/8$.

The coupling constant has dimension of $|p|^{-1}$; taking $|p|$ as a running scale we define a dimensionless coupling 
constant $$g'=|p|g=\frac{|p|\lambda_0}{1-\lambda_0[B(T,3]+A(T,3)|p|}$$ and 
the beta-function,
\begin{equation}
\beta(g')=|p|\frac{\partial\,g'}{\partial |p|}; 
\label{beta}
\end{equation}
we  easily see that the condition of a non-trivial infrared stable fixed point 
is fullfiled by the solution 
\begin{equation} 
g'_{\star}=\frac{1}{A(3,T_c)}.
\label{fixed}
\end{equation}

As we have already stated in this note, the GN model may be seen
as an effective theory for QCD, with an arbitrary zero-temperature
coupling constant. Let us then estimate a specific value for the
critical temperature,  choosing the mass of the Gross-Neveu
fermion to be the effective quark mass of the proton~\cite{oct201},
$m_{0}\approx 330\,{\rm{MeV}}$. Consider  a strong  coupling
regime  characterized by large values of $\lambda_0$, up to the limit $ \lambda_0
\rightarrow \infty$ (see Fig.~\ref{figura3}). For these values (for instance for $\lambda_0\approx 16\pi/m_0=0.15\,{\rm{MeV}}^{-1}$, corresponding to $\lambda=8$ in Fig.~\ref{figura3}), we find a critical temperature, $T_c$, of the order of $T_{c}\sim
0.87m_{0}\approx 288\,{\rm{MeV}}$.


\section{Concluding remarks}
In short, we have answered positively the question whether there
is a phase transition in the $N=1$ massive GN model; i.e.
yes, there is a second-order phase transition.
 This conclusion is
based in two results. From the non-perturbative
analysis of the four-point function, a critical region can be
established. Thereby the analysis of the free-energy provides the
nature of the transition, as being of second-order. It is
important to add that this procedure is strikingly similar in
character to the method for finding critical temperature for the
second-order phase transition in BCS model of superconductivity. Also, 
we have shown that we can define a $beta$-function, which has a non-trivial infrared fixed 
point, thus reinforcing the conclusion that the transition is a second-order one. 

Considering a regime of strong coupling as defined above, and  the mass of
the GN fermion as the effective quark mass, we find a
critical temperature of the order of
 $T_{c}\simeq 288 MeV$. It is interesting to observe that this value
 is of the same order of magnitude as the
estimated temperature $\sim 200 \rm{MeV}$ for the quark
deconfinement transition, obtained from lattice calculations. The
phase-transition here may be then associated with the transition
from a hadronic state to a quark-gluon plasma.

 This has been possible by employing the {\it massive} GN model in three dimensions, which has been shown to exist and constructed for the first time in~\cite{Jarrao}. In previous works devoted to get insights on the behaviour of hadronic matter~\cite{gn2,gn3,gn5,gn6,gn7,gn8,gn9,gn10,gn11,gn12,gn13,gn14,gn16,gn17,gn18},  the massless GN model, in its version with a large number of components, has been often employed. Using the one-component massive GN model in three dimensions and taking the fermion mass as a physical parameter (the effective quark mass), we have been able using {\it analytical means}, to determine  the transition temperature. Moreover this temperature is found to fall into a range of values compatible with the  transition temperature for hadronization, estimated from lattice simulations. A rigorous study involving aditional  aspects of this transition, including determination of critical exponents, is left for future work. 

Finally, we would like to add some comments about the  suitability of the GN model to fit QCD.  We have in mind the Nambu-Jona-Lasinio (NJL) model which, particularly in its four-dimensional version is closer to QCD than the GN model. Many studies on the phase structure of the two and four-dimensional NJL model have been performed, which includes articles by some of us (see~\cite{luc1,luc2} and references therein). A relevant aspect of the massive three dimensional GN model, as we have employed in this article, is that the critical temperature can be obtained analytically in the context of a well known phenomenological approach to phase transitions. This is much harder to be studied with the NJL model; it will the subject of future work.

{\bf Acknowledgments}

 This work  received
partial financial support from CAPES, CNPq, FAPERJ (Brazil) and NSERC
(Canada).

\end{document}